\pdfoutput=1
\documentclass [12pt]{article}

\usepackage[hyperindex]{hyperref}

\def\maj#1{\ifmmode\mbox{\usefont{U}{msb}{m}{n}#1}\else{\usefont{U}{msb}{m}{n}#1}\fi}
\def\v#1{\mathbf{#1}}
\usepackage{graphics,epsfig,graphicx}
\usepackage{color}
\makeatletter
\@addtoreset{equation}{section}
\makeatother

\begin{document}

\title{\textbf{Exciton many-body effects through infinite series of  
composite-exciton operators}}
\author{M. Combescot and O. Betbeder-Matibet
\\ \small{\textit{Institut des NanoSciences de Paris,}}\\
\small{\textit{Universit\'e Pierre et Marie Curie, 
CNRS,}}\\
\small{\textit{Campus Boucicaut, 140 rue de Lourmel, 75015 Paris}}}
\date{}
\maketitle

\begin{abstract}
We revisit the approach proposed by Mukamel and coworkers to describe
interacting excitons through infinite series of composite-boson operators
for both, the system Hamiltonian and the exciton commutator --- which,
in this approach, is properly kept different from its elementary boson value. Instead of free electron-hole operators, as used by Mukamel's group, we here work with composite-exciton operators which are physically relevant operators for excited semiconductors.
This allows us to get \emph{all} terms of these infinite series
explicitly, the first terms of each series agreeing with the ones obtained by Mukamel's group when written with electron-hole pairs. All these terms nicely read in terms of Pauli and interaction scatterings of the
composite-exciton many-body theory we have recently proposed. However, even if knowledge of these infinite series now allows to tackle $N$-body problems, not just 2-body problems like third order nonlinear susceptibility $\chi^{(3)}$, the necessary handling of these two infinite series makes this approach far more complicated than the one we have developed and which barely relies on just four commutators.
\end{abstract}


\newpage

\section{Introduction}

Most particles known as bosons, are composite particles made of 
even number of fermions. Proper treatment of the underlying Pauli
exclusion principle between fermionic components of these particles
has been a longstanding problem for decades [1]. Because many-body theories for
quantum particles were, up to our work, valid for elementary
particles only [2], sophisticated ``bosonization'' procedures [3,4] have been
proposed to replace composite bosons by elementary bosons. These
elementary bosons then interact through effective scatterings
constructed on interactions which exist between their fermionic components,
but dressed by ``appropriate'' fermion exchanges [5]. Although quite popular
due to the fact that they allow calculations on problems otherwise
unsolvable through known procedures, such bosonizations have
an intrinsic major failure linked to the fact that, by replacing two free
fermions by one boson, we strongly reduce degrees of freedom of the
system. This shows up through the fact that, while closure relation
for $N$ elementary bosons is
\begin{equation}
\bar{I}=\frac{1}{N!}\sum\bar{B}_{i_1}^\dag\ldots\bar{B}_{i_N}^\dag|v
\rangle\,\langle v|\bar{B}_{i_N}\ldots\bar{B}_{i_1}\ ,
\end{equation}
with $[\bar{B}_i,\bar{B}_j]=\delta_{ij}$, the one for $N$ composite
bosons made of two free fermions reads [6]
\begin{equation}
I=\frac{1}{(N!)^2}\sum B_{i_1}^\dag\ldots B_{i_N}^\dag|v
\rangle\,\langle v|B_{i_N}\ldots B_{i_1}\ .
\end{equation}
The huge prefactor change from $N!$ to $(N!)^2$ makes all sum rules for elementary and composite bosons, based on
this closure relation, irretrievably different whatever are effective
scatterings produced by bosonization procedures. And indeed, through this
prefactor difference in closure relations, we have explained [6] the factor 1/2 difference in the link between lifetime and
sum of transition rates that we had found [7] for composite and bosonized excitons.

Besides bosonization, very few other approaches to interacting composite
bosons have been proposed. In the late 60's, M. Girardeau [8] suggested to
introduce a set of ``ideal atom operators'' in addition to fermionic
operators for electrons and protons. These operators, which are bosonic
by construction, represent all bound states of one atom, but \emph{not} its
extended states. They are forced into the problem through a so-called
Fock-Tani unitary transformation which, in an exact way, transforms
\emph{one} exact atom bound state into one ideal-atom state.
Unfortunately, this nicely simple result does not hold for $N$-atom states with $N\geq 2$, the
procedure turning quite complicated very fast. This is why, although not
advocated by Girardeau, we can be tempted by using his procedure as a bosonization procedure, \emph{i.e.}, by only keeping ideal-atom operators in transformed states and
transformed Hamiltonian. We have however shown [9] that, with such a
reduction, the obtained results for a few relevant physical quantities are at odd from
the correct ones, even for the sign. The idea to add to fermionic
operators for electrons and protons,
a set of bosonic operators for atomic bound states, is in fact
rather awkward because fermionic operators form a complete set in
themselves; so that Girardeau artificially introduces an overcomplete set of operators in
a problem already complex, this overcompleteness being obviously difficult to
handle properly. Precise comparison of Girardeau's procedure
with the composite-boson many-body theory we have constructed, can be found
in reference [9].

Another approach, still currently used [10-12], has been proposed by
Mukamel and coworkers in the 90's. It is based on the fully correct idea that the
system Hamiltonian, when acting on fermion pairs, can be replaced by an
infinite series of pair operators. In this approach, the fact that pairs
of fermions differ from elementary bosons is kept exactly through
commutators of pair operators which are also written as infinite series. The pair-operators used by
Mukamel and coworkers are products of free fermion
operators. However, as these are not physically relevant pair operators for problems dealing with
excitons, their calculations turn out very
complicated. This is probably why they have only derived the first term of the Hamiltonian and pair-commutator series. This thus makes their
results of possible use for problems restricted to two excitons only.
And indeed, using them, they have successfully calculated [12] the third order
susceptibility $\chi^{(3)}$ which results from interactions of two
unabsorbed photons through their coupling to two virtual excitons.

In this paper, we follow Mukamel and coworkers' idea, but with exciton operators $B_i^\dag$
instead of products of free-electron and free-hole operators $a_{\v k_e}^\dag b_{\v k_h}^\dag$, these
exciton operators being the ones which create one-electron-hole-pair eigenstates of the system Hamiltonian,
\begin{equation}
(H-E_i)B_i^\dag|v\rangle=0\ .
\end{equation}
Thanks to the closure relation for $N$ composite
excitons, Eq.(1.2), it is easy to derive all terms of the series for the
composite-boson commutator and for the system Hamiltonian in an exact way. As expected, prefactors in these infinite series read in
terms of the two key parameters of the composite-boson many-body physics,
namely Pauli scatterings for fermion exchanges in the absence of
fermion interaction, and interaction scatterings for fermion
interactions in the absence of fermion exchange.

However, even with these two infinite series at hand explicitly, so that
problems dealing with many-body effects between $N$ excitons could now be 
tackled, this approach turns out to be definitely far
more complicated than the composite-boson many-body theory we have recently
constructed [13,14]. Indeed, in this new theory, calculations dealing with many-body effects between any number
$N$ of excitons simply reduce to performing a set of commutations between exciton
operators, according to two commutators for fermion exchanges (see Eq.(5) in ref. [15] or Eq. (14) in ref. [13]), namely,
\begin{equation}
[B_m,B_i^{\dag N}]=NB_i^{\dag N-1}(\delta_{m,i}-D_{mi})-N(N-1)B_i^{\dag
N-2}\sum_n\lambda\left(^{n\ \,i}_{m\ i}\right)B_n^\dag\ ,
\end{equation}
\begin{equation}
[D_{mi},B_j^{\dag N}]=NB_j^{\dag N-1}\sum_n\left\{\lambda\left(^{n\
\,j}_{m\ i}\right)+\lambda\left(^{m\ j}_{n\ \,i}\right)\right\}B_n^\dag\ ,
\end{equation}
and two commutators for fermion interactions (see Eq.(5) in ref. [16 ] or Eq.(13) in ref.[13]), namely,
\begin{equation}
[H,B_i^{\dag N}]=NB_i^{\dag N-1}(E_iB_i^\dag+V_i^\dag)+\frac{N(N-1)}{2}
B_i^{\dag N-1}\sum_{m,n}\xi\left(^{n\ \,i}_{m\ i}\right)B_m^\dag
B_n^\dag\ ,
\end{equation}
\begin{equation}
[V_i^\dag,B_j^{\dag N}]=NB_j^{\dag N-1}\sum_{m,n}\xi\left(^{n\ \,j}_{m\ i}
\right)B_m^\dag B_n^\dag\ .
\end{equation}
In these equations, $D_{mi}$ is the exciton ``deviation-from-boson operator'' defined through
Eq.(1.4) taken for $N=1$, namely,
\begin{equation}
D_{mi}=\delta_{m,i}-[B_m,B_i^\dag]\ ,
\end{equation}
while Pauli scattering $\lambda\left(^{n\ \,j}_{m\ i}\right)$ of two
``in'' excitons $(i,j)$ towards two ``out'' excitons $(m,n)$ follows from
Eq.(1.5) taken for $N=1$ (also see Eq.(4) in ref.[17]), namely,
\begin{equation}
[D_{mi},B_j^\dag]=\sum_n\left\{\lambda\left(^{n\ \,j}_{m\ i}\right)+
\lambda\left(^{m\ j}_{n\ \,i}\right)\right\}B_n^\dag\ .
\end{equation}

In the same way, ``creation potential'' $V_i^\dag$ of exciton $i$
and interaction scattering $\xi\left(^{n\ \,j}_{m\ i}\right)$ follow
from Eqs.(1.6) and (1.7) taken for $N=1$ (also see Eq.(3) in ref.[17]), namely,
\begin{equation}
[H,B_i^\dag]=E_iB_i^\dag+V_i^\dag\ ,
\end{equation}
\begin{equation}
[V_i^\dag,B_j^\dag]=\sum_{m,n}\xi\left(^{n\ \,i}_{m\ i}\right)B_m^\dag
B_n^\dag\ .
\end{equation}

Let us note that Eq.(1.8) which basically says that particles are not elementary but composite bosons, was known for quite a long time [1,18]. Equation (1.10) is more recent. It was introduced by one of us in her theory of exciton optical Stark effect [19,20]. On the contrary, Eqs.(1.9) and (1.11) which allow to reach the two elementary scatterings of two excitons, namely, $\lambda\left(^{n\ \,j}_{m\ i}\right)$ and $\xi\left(^{n\ \,j}_{m\ i}\right)$, are fundamentally new. They are the keys of our composite-boson many-body theory [17].

In this paper, we are going to use these four commutators to write the system Hamiltonian $H$
and the deviation-from-boson operator $D_{mi}$ as infinite series of
exciton operators $B_i^\dag$. This will allow us to generate physically relevant
prefactors for these series. They are found to read in terms of
exciton energies $E_i$, interaction scattering of two excitons
$\xi\left(^{n\ \,j}_{m\ i}\right)$, and the following sum of
Pauli scatterings,
\begin{eqnarray}
\Lambda_{mi}(n,j)&=&\lambda\left(^{n\ \,j}_{m\ i}\right)+
\lambda\left(^{m\ j}_{n\ \,i}\right)\nonumber\\
&=&\lambda_e\left(^{n\ \,j}_{m\ i}\right)+\lambda_h\left(^{n\ \,j}_{m\
i}\right)\ .
\end{eqnarray}
$\Lambda_{mi}(n,j)$, shown in Fig.1, corresponds to processes in which
excitons $(i,j)$ exchange either a hole or an electron, excitons
$(m,i)$ having same electron in $\lambda_e\equiv\lambda$, while they have same hole in
$\lambda_h$: Due to electron-hole symmetry, it is quite reasonable to
find these two processes on the same footing, in the $\Lambda_{mi}(n,j)$ factor.

\section{Deviation-from-boson operator}

Let us start with deviation-from-boson operator $D_{mi}$ defined in
Eq.(1.8). Since both, $D_{mi}$ and the product of exciton operators $B_m^\dag B_i$ conserve number of pairs, we can look for $D_{mi}$  as
\begin{equation}
D_{mi}=\sum_{n=1}^{\infty}D_{mi}^{(n)}\ ,
\end{equation}
where the most general form for $D_{mi}^{(n)}$ acting in the subspace made of states having $p\geq n$ pairs, can be taken as
\begin{equation}
D_{mi}^{(n)}=\sum_{\{r\}}d_{mi}^{(n)}(r'_1,\ldots,r'_n;r_1,\ldots,r_n)
B_{r'_1}^\dag\ldots B_{r'_n}^\dag B_{r_1}\ldots B_{r_n}\ .
\end{equation}
We get this series by enforcing it to be such that, when acting on
any $N$-exciton state $|\psi_N\rangle$ linear combination of 
$B_{j_1}^\dag\ldots B_{j_N}^\dag|v\rangle$, it gives the same result as
the original operator $D_{mi}$, namely,
\begin{equation}
D_{mi}|\psi_N\rangle=\sum_{n=1}^N D_{mi}^{(n)}|\psi_N\rangle\ ,
\end{equation}
for any $N$-pair state $|\psi_N\rangle$. We are going
to derive the various operators $D_{mi}^{(n)}$ by iteration,
starting from $n=1$, as we now show.

Before going further, it is of importance to note that, due to carrier
exchanges between two excitons, we do have (see Eq.(5) in ref.Ê[17])
\begin{equation}
B_i^\dag B_j^\dag =-\sum_{m,n}\lambda\left(^{n\ \,j}_{m\ i}\right)B_m^\dag
B_n^\dag\ .
\end{equation}
This equation, which comes from the two ways to construct two excitons out of two
electron-hole pairs, shows that $N$-exciton states $|\psi_N\rangle$ for
$N\geq 2$, as well as operators like $D_{mi}^{(n)}$ for $n\geq 2$ can be
written in many different ways, these various forms being related through
the replacement of any $B^\dag B^\dag$ by a sum of $B^\dag B^\dag$
according to Eq.(2.4): Just as $B_{i_1}^\dag\ldots B_{i_N}^\dag|v\rangle$ states form an
overcomplete set for $N$-pair states, $B_i^\dag$'s form an
overcomplete set of operators. This, in particular, allows us to guess that, among the
various possible forms of $D_{mi}^{(n)}$, the one which has a physically
relevant meaning, is most probably the simplest one. We will come back to this
fundamental indetermination, linked to exciton composite nature, at the end of this section.

\subsection{Derivation of $D_{mi}^{(1)}$}

Let us first consider a one-exciton state $|\psi_1\rangle$. By
inserting closure relation for one-exciton subspace , \emph{i.e.}, Eq.(1.2) taken for $N=1$, in front of this state, we find
\begin{equation}
D_{mi}|\psi_1\rangle=\sum_{r_1}D_{mi}B_{r_1}^\dag|v\rangle\langle v|
B_{r_1}|\psi_1\rangle\ .
\end{equation}
As $D_{mi}B_{r_1}^\dag|v\rangle=[D_{mi},B_{r_1}^\dag]|v\rangle$ since
$D_{mi}|v\rangle=0$ which follows from Eq.(1.8) acting on vacuum, we get
from Eqs.(1.9,12)
\begin{equation}
D_{mi}|\psi_1\rangle=\sum_{r'_1,r_1}\Lambda_{mi}(r'_1,r_1)B_{r'_1}^\dag |v
\rangle\langle v|B_{r_1}|\psi_1\rangle\ ,
\end{equation}
where $\Lambda_{mi}(n,j)$ is the combination of Pauli scatterings introduced in Eq.(1.12).

We then note that projector $|v\rangle\langle v|$ can be removed from
this equation since state $B_{r_1}|\psi_1\rangle$ has zero pair
while identity operator reduces to $|v\rangle\langle v|$ for such a state. Consequently, Eq.(2.6) also reads
\begin{equation}
D_{mi}|\psi_1\rangle=\sum_{r'_1,r_1}\Lambda_{mi}(r'_1,r_1)B_{r'_1}^\dag B_{r_1}|
\psi_1\rangle\ .
\end{equation}
Since this equation is valid for any state $|\psi_1\rangle$, we readily
find that operator $D_{mi}^{(1)}$ such that $D_{mi}|\psi_1\rangle=
D_{mi}^{(1)}|\psi_1\rangle$ can be taken as
\begin{equation}
D_{mi}^{(1)}=\sum_{r'_1,r_1}\Lambda_{mi}(r'_1,r_1)B_{r'_1}^\dag B_{r_1}\ ,
\end{equation}
with $\Lambda_{mi}(r'_1,r_1)$ given in Eq.(1.12). This result is the same as the one given by Mukamel and coworkers (see Eqs. (11-13) of Ref.[12]).

\subsection{Derivation of $D_{mi}^{(2)}$}

We now consider two-exciton state $|\psi_2\rangle$. By inserting closure
relation, Eq.(1.2), for two-exciton subspace, in front of $|\psi_2\rangle$, we get
\begin{equation}
D_{mi}|\psi_2\rangle=\left(\frac{1}{2!}\right)^2\sum_{r_1,r_2}D_{mi}
B_{r_1}^\dag B_{r_2}^\dag|v\rangle\langle v|B_{r_2}B_{r_1}|\psi_2\rangle\
.
\end{equation}
To go further, we note that, due to Eqs.(1.9) and (1.12),
\begin{eqnarray}
D_{mi}B_{r_1}^\dag B_{r_2}^\dag|v\rangle&=& \left([D_{mi},B_{r_1}^\dag]+
B_{r_1}^\dag D_{mi}\right)B_{r_2}^\dag|v\rangle\nonumber\\
&=&\sum_{r'}B_{r'}^\dag\left(\Lambda_{mi}(r',r_1)B_{r_2}^\dag+\Lambda_{mi}(r',r_2)
B_{r_1}^\dag\right)|v\rangle\ .
\end{eqnarray}
We then insert this result into Eq.(2.9) and relabel bold indices.
By noting that projector $|v\rangle\langle v|$ can again be removed
from this equation since $B_{r_2}B_{r_1}|\psi_2\rangle$  also has
zero pair, we end with
\begin{equation}
D_{mi}|\psi_2\rangle=\frac{2}{(2!)^2}\sum_{r'_1,r_1,r_2}\Lambda_{mi}(r'_1,r_1)
B_{r'_1}^\dag B_{r_2}^\dag B_{r_2}B_{r_1}|\psi_2\rangle\ .
\end{equation}
If we now turn to $D_{mi}^{(1)}$ acting on $|\psi_2\rangle$, we note that
$B_{r_1}|\psi_2\rangle$ has one pair so that, if we insert
closure relation for one-pair subspace in front of this state, we find
\begin{equation}
D_{mi}^{(1)}|\psi_2\rangle=\sum_{r'_1,r_1}\Lambda_{mi}(r'_1,r_1)B_{r'_1}^\dag
\left[\sum_{r_2}B_{r_2}^\dag|v\rangle\langle v|B_{r_2}\right]B_{r_1}|\psi_2\rangle\ .
\end{equation}
We can again remove projector $|v\rangle\langle v|$ from this
equation since $B_{r_2}B_{r_1}|\psi_2\rangle$ has zero pair. This
readily shows that operator $D_{mi}^{(2)}$ such that $D_{mi}|\psi_2
\rangle=\left(D_{mi}^{(1)}+D_{mi}^{(2)}\right)|\psi_2\rangle$ can be
taken as
\begin{equation}
D_{mi}^{(2)}=-\frac{1}{2}\sum_{r'_1,r_1,r_2}\Lambda_{mi}(r'_1,r_1)B_{r'_1}^\dag
B_{r_2} ^\dag B_{r_2}B_{r_1}\ .
\end{equation}

\subsection{Derivation of $D_{mi}^{(3)}$}

To better grasp how $D_{mi}^{(n)}$ can be constructed by iteration,
let us calculate one more $D_{mi}^{(n)}$ explicitly. We consider
three-pair state $|\psi_3\rangle$ and inject in front of it, closure
relation for three-pair subspace. This leads to
\begin{equation}
D_{mi}|\psi_3\rangle=\frac{1}{(3!)^2}\sum_{r_1,r_2,r_3}D_{mi}B_{r_1}^\dag
B_{r_2}^\dag B_{r_3}^\dag|v\rangle\langle v|B_{r_3}B_{r_2}B_{r_1}|\psi_3
\rangle\ .
\end{equation}
To go further, we do like for Eq.(2.10) and use commutator $[D_{mi},
B_r^\dag]$ given in Eq.(1.9). This leads to
\begin{eqnarray}
D_{mi}B_{r_1}^\dag B_{r_2}^\dag B_{r_3}^\dag|v\rangle=\sum_{r'}\left\{
\Lambda_{mi}(r',r_1)B_{r_2}^\dag B_{r_3}^\dag+\Lambda_{mi}(r',r_2)B_{r_1}^\dag
B_{r_3}^\dag\right.\nonumber\\
\left.+\Lambda_{mi}(r',r_3)B_{r_1}^\dag B_{r_2}^\dag\right\}|v\rangle\ .
\end{eqnarray}
We then inject this result into Eq.(2.14), relabel bold indices and
remove projector $|v\rangle\langle v|$. This leads to
\begin{equation}
D_{mi}|\psi_3\rangle=\frac{3}{(3!)^2}\sum_{r'_1,\{r\}}\Lambda_{mi}(r'_1,r_1)
B_{r'_1}^\dag B_{r_2}^\dag B_{r_3}^\dag
B_{r_3}B_{r_2}B_{r_1}|\psi_3\rangle\ .
\end{equation}

We now turn to $D_{mi}^{(1)}|\psi_3\rangle$. Since $B_{r_1}|\psi_3\rangle$
has two pairs, we get, by using closure relation for 
two-pair subspace,
\begin{equation}
D_{mi}^{(1)}|\psi_3\rangle =\left(\frac{1}{2!}\right)^2\sum_{r'_1,\{r\}}
\Lambda_{mi}(r'_1,r_1)
B_{r'_1}^\dag B_{r_2}^\dag B_{r_3}^\dag\v\rangle\langle v|
B_{r_3}B_{r_2}B_{r_1}|\psi_3\rangle\ .
\end{equation}
We do the same for $D_{mi}^{(2)}|\psi_3\rangle$ in which $B_{r_1}B_{r_2}|
\psi_3\rangle$ has one pair. By collecting all terms, we see
that operator $D_{mi}^{(3)}$ such that $D_{mi}|\psi_3\rangle=\left(
D_{mi}^{(1)}+D_{mi}^{(2)}+D_{mi}^{(3)}\right)|\psi_3\rangle$ can be taken
as
\begin{equation}
D_{mi}^{(3)}=\frac{1}{3}\sum_{r'_1,\{r\}}\Lambda_{mi}(r'_1,r_1)B_{r'_1}^\dag
B_{r_2}^\dag B_{r_3}^\dag B_{r_3}B_{r_2}B_{r_1}\ .
\end{equation}

\subsection{Derivation of $D_{mi}^{(n)}$}

The above results lead us to think that operator $D_{mi}^{(n)}$ can
be written as
\begin{equation}
D_{mi}^{(n)}=\gamma_n\sum_{r'_1,\{r\}}\Lambda_{mi}(r'_1,r_1)B_{r'_1}^\dag B_{r_2}
^\dag\ldots B_{r_n}^\dag B_{r_n}\ldots B_{r_2}B_{r_1}\ ,
\end{equation}
where $\gamma_n$ is a numerical prefactor which, in spite of its values for $n=(1,2,3)$, does not reduce to $(-1)^{n-1}/n$.

To determine $\gamma_n$, we look for the recursion relation it obeys. To get this recursion relation, we
follow the procedure we have used for $n\leq 3$, namely, we insert  closure relation for
$N$-pair subspace in front of state $|\psi_N\rangle$. This leads to
\begin{equation}
D_{mi}|\psi_N\rangle=\left(\frac{1}{N!}\right)^2\sum_{\{r\}}D_{mi}
B_{r_1}^\dag\ldots B_{r_N}^\dag|v\rangle\langle v|B_{r_N}\ldots B_{r_1}|
\psi_N\rangle\ .
\end{equation}
We then calculate $D_{mi}B_{r_1}^\dag\ldots B_{r_N}^\dag|v\rangle$ using
commutator (1.9);  we relabel bold indices and remove projector
$|v\rangle\langle v|$. This gives
\begin{equation}
D_{mi}|\psi_N\rangle=\frac{N}{(N!)^2}\sum_{r'_1,\{r\}}\Lambda_{mi}(r'_1,r_1)
B_{r'_1}^\dag B_{r_2}^\dag\ldots B_{r_N}^\dag B_{r_N}\ldots B_{r_2}B_{r_1}
|\psi_N\rangle\ .
\end{equation}
We then turn to $D_{mi}^{(n)}$ acting on $|\psi_N\rangle$ for $n<N$. Since
state $B_{r_1}\ldots B_{r_n}|\psi_N\rangle$ has $(N-n)$ pairs, 
closure relation for this subspace leads to
\begin{eqnarray}
D_{mi}^{(n)}|\psi_N\rangle=\left(\frac{1}{(N-n)!}\right)^2\gamma_n
\sum_{r'_1,\{r\}}\Lambda_{mi}(r'_1,r_1)B_{r'_1}^\dag B_{r_2}^\dag\ldots B_{r_n}
B_{r_{n+1}}^\dag\ldots B_{r_N}^\dag\nonumber\\
\times\ B_{r_N}\ldots B_{r_{n+1}}B_{r_n}\ldots B_{r_1}|\psi_N\rangle\ .
\end{eqnarray}
By inserting these results into Eq.(2.3), it is easy to show that the form
Eq.(2.19) for $D_{mi}^{(n)}$ is indeed valid provided that $\gamma_n$'s are linked by
\begin{equation}
\gamma_N=\frac{N}{(N!)^2}-\sum_{n=1}^{N-1}\frac{\gamma_n}{[(N-n)!]^2}\ ,
\end{equation}
with $\gamma_1=1$. From this equation, it is easy to show that the first $\gamma_n$'s are 
\begin{eqnarray}
\gamma_2&=&-1/2\nonumber\\
\gamma_3&=&1/3\nonumber\\
\gamma_4&=&-11/48\nonumber\\
\gamma_5&=&11/120\ ,
\end{eqnarray}
and so on\ldots, with 
$\gamma_N$ going to
zero with increasing $N$.

\subsection{Other forms of $D_{mi}^{(n)}$}

As said at the beginning of this section, composite-boson operators
$B_i^\dag$ form an overcomplete set to describe electron-hole pairs.
This is why any given operator acting in $N$-pair subspace with $N\geq 2$, when
written in terms of these $B_i^\dag$'s, can appear through different
expressions. Indeed, due to Eq.(2.4), it is possible to rewrite $B_{r'_1}^\dag B_{r_2}^\dag$ in Eq.(2.19) as
\begin{equation}
B_{r'_1}^\dag B_{r_2}^\dag=-\sum_{r''_1,r'_2}\lambda\left(^{r''_1\ r_2}
_{r'_2\ \,r'_1}\right)B_{r''_1}^\dag B_{r'_2}^\dag\ ,
\end{equation}
since $B_m^\dag B_n^\dag=B_n^\dag B_m^\dag$. We then note that
\begin{equation}
\sum_{r'_1}\lambda\left(^{r''_1\ r_2}_{r'_2\ \,r'_1}\right)\lambda\left(^{r'_1\ r_1}
_{m\ \,i}\right)
=\lambda_3\left(\begin{array}{ll}r'_2&r_2\\r''_1&r_1\\m&i\end{array}
\right)\ ,
\end{equation}
where, according to Fig.2(a), $\lambda_3$ is just the exchange scattering between three excitons $(i,r_1,r_2)$. This allows us to replace
the first factor of
$D_{mi}^{(n)}$ in Eq.(2.19) by
\begin{equation}
\sum_{r'_1}\Lambda_{mi}(r'_1,r_1)B_{r'_1}^\dag B_{r_2}^\dag=-\sum_{r''_1,r'_2}
\Lambda_{mi}\left(^{r'_2\ r_2}_{r''_1\ r_1}\right)B_{r''_1}^\dag B_{r'_2}
^\dag\ .
\end{equation}
While prefactor $\Lambda_{mi}(r'_1,r_1)$ corresponds to carrier exchanges between two excitons $(i,r_1)$ leading to $(m,r'_1)$ with excitons $m$ and $i$ having either same electron or same hole,  prefactor $\Lambda_{mi}\left(^{r'_2\ r_2}_{r''_1\ r_1}\right)$
corresponds to carrier exchanges between excitons
$(i,r_1,r_2)$ leading to $(m,r''_1,r'_2)$, with excitons $m$ and $i$ also having either
same electron or same hole (see Fig.2(b)).

If we keep doing this procedure for $B_{r'_2}^\dag B_{r_3}^\dag$ with
$B_{r'_2}^\dag$ relabelled as $B_{r'_1}^\dag$, and so on \ldots, we end
with $D_{mi}^{(n)}$ written in a quite symmetrical form, although far
more complicated than Eq.(2.19), namely,
\begin{equation}
D_{mi}^{(n)}=(-1)^{n-1}\gamma_n\sum_{\{r'\},\{r\}}\Lambda_{mi}\left(
\begin{array}{ll}r'_n&r_n\\.&.\\.&.\\r'_1&r_1\end{array}\right)
B_{r'_1}^\dag\ldots B_{r'_n}^\dag B_{r_n}\ldots B_{r_1}\ ,
\end{equation}
where the prefactor corresponds to carrier exchanges between $(n+1)$ excitons
$(i,r_1,\ldots,r_n)$ leading to $(m,r'_1,\ldots,r'_n)$ in which excitons $m$ and
$i$ either have same electron or same hole (see Fig.2(c)).

\section{System Hamiltonian}

Let us now turn to the system Hamiltonian originally written in terms of
fermionic operators for free electrons and free holes. It contains 
kinetic electron and hole contributions. It also contains Coulomb interaction
between electrons, between holes and between electrons and holes. It is actually quite easy to write
the electron-hole part of this Hamiltonian in terms of excitons.  Indeed, by using
the link between exciton operators and free-electron and free-hole
operators, namely,
\begin{equation}
B_i^\dag=\sum_{\v k_e,\v k_h}a_{\v k_e}^\dag b_{\v k_h}^\dag\langle\v k_h,
\v k_e|i\rangle\ ,
\end{equation}
\begin{equation}
a_{\v k_e}^\dag b_{\v k_h}^\dag=\sum_iB_i^\dag\langle i|\v k_e,\v
k_h\rangle\ ,
\end{equation}
where $\langle\v k_h,\v k_e | i\rangle$ is $i$ exciton wave function in momentum space,
we readily find electron-hole Coulomb interaction as
\begin{eqnarray}
V_{eh}&=&-\sum_{\v q,\v k_e,\v k_h}V_{\v q}a_{\v k_e+\v q}^\dag b_{\v
k_h-\v q}^\dag b_{\v k_h} a_{\v k_e}\nonumber\\
&=& -\sum_{i,j}B_i^\dag B_j\sum_{\v q,\v k_e,\v k_h}V_{\v q}
\langle i|\v k_e+\v q,\v k_h-\v q\rangle\langle\v k_h,\v k_e|j\rangle\ .
\end{eqnarray}
On the contrary, this cannot be done for other parts of the
Hamiltonian, namely, kinetic energy terms in $a^\dag a$ and $b^\dag b$ and
electron-electron and hole-hole Coulomb terms in $a^\dag a^\dag aa$ and
$b^\dag b^\dag bb$. Nevertheless, since both operator $H$ and product of exciton operators $B_m^\dag B_i$, conserve the
number of electron-hole pairs, it is \emph{a priori} possible to write $H$ as
\begin{equation}
H=\sum_{n=1}^{\infty} H^{(n)}\ ,
\end{equation}
\begin{equation}
H^{(n)}=\sum_{\{r\},\{r'\}}h^{(n)}(r'_1,\ldots,r'_n;r_1,\ldots,r_n)
B_{r'_1}^\dag\ldots B_{r'_n}^\dag B_{r_n}\ldots B_{r_1}\ ,
\end{equation}
so that $H^{(n)}$ acts on states having $p\geq n$ pairs. This series is
determined by enforcing
\begin{equation}
H|\psi_N\rangle=\sum_{n=1}^N H^{(n)}|\psi_N\rangle\ ,
\end{equation}
for any state $|\psi_N\rangle$ having $N$ electron-hole pairs. Here
again, $H^{(n)}$ for $n\geq 2$ is expected to have various forms since
any $B^\dag B^\dag$ can be replaced by sum of $B^\dag B^\dag$,
according to Eq.(2.4). To get the various terms of $H^{(n)}$
expansion, we are again going to extensively use closure relation
(1.2) for $N$-pair states. This will allow us to get one of these
possible forms of $H$ quite easily.

\subsection{Derivation of $H^{(1)}$}

To get $H^{(1)}$, we insert closure relation for one-pair states in
front of $|\psi_1\rangle$ in $H|\psi_1\rangle$. This leads to
\begin{equation}
H|\psi_1\rangle=\sum_{r_1}HB_{r_1}^\dag|v\rangle\langle v|B_{r_1}|\psi_1
\rangle\ .
\end{equation}
We first replace $HB_{r_1}^\dag|v\rangle$ by
$E_{r_1}B_{r_1}^\dag|v\rangle$ for exciton operators create
one-pair eigenstates of the system. We then note that $B_{r_1}|\psi_1
\rangle$ has zero pair, so that we can remove projector
$|v\rangle\langle v|$ from this equation. This leads to
\begin{equation}
H|\psi_1\rangle=\sum_{r_1}E_{r_1}B_{r_1}^\dag B_{r_1}|\psi_1\rangle\ .
\end{equation}
Since $H^{(1)}|\psi_1\rangle$ must be equal to $H|\psi_1\rangle$ for any one-pair state $|\psi_1\rangle$, we readily see that
$H^{(1)}$ can be identified with
\begin{equation}
H^{(1)}=\sum_{r_1}E_{r_1}B_{r_1}^\dag B_{r_1}\ .
\end{equation}

\subsection{Derivation of $H^{(2)}$}

We now turn to two-pair subspace. By inserting closure relation
for two-pair states in front of $|\psi_2\rangle$, we find
\begin{equation}
H|\psi_2\rangle=\left(\frac{1}{2!}\right)^2\sum_{r_1,r_2}HB_{r_1}^\dag
B_{r_2}^\dag|v\rangle\langle v|B_{r_2}B_{r_1}|\psi_2\rangle\ .
\end{equation}
We then use Eqs.(1.10,11) to find
\begin{eqnarray}
HB_{r_1}^\dag B_{r_2}^\dag|v\rangle&=&(B_{r_1}^\dag H+E_{r_1}B_{r_1}^\dag
+V_{r_1}^\dag)B_{r_2}^\dag|v\rangle\nonumber\\
&=& (E_{r_1}+E_{r_2})B_{r_1}^\dag B_{r_2}^\dag|v\rangle+\sum_{r'_1,r'_2}
\xi\left(^{r'_2\ r_2}_{r'_1\ r_1}\right)B_{r'_1}^\dag B_{r'_2}^\dag|v
\rangle\ .
\end{eqnarray}
If we insert this result into Eq.(3.10), relabel bold indices and
remove projector $|v\rangle\langle v|$, we end with
\begin{equation}
H|\psi_2\rangle=\left(\frac{1}{2}\sum_{r_1,r_2}E_{r_1}B_{r_1}^\dag B_{r_2}
^\dag B_{r_2} B_{r_1}+\frac{1}{4}\sum_{r_1,r_2,r'_1,r'_2}\xi
\left(^{r'_2\ r_2}_{r'_1\ r_1}\right)B_{r'_1}^\dag
B_{r'_2}B_{r_2}B_{r_1}\right)|\psi_2\rangle\ .
\end{equation}

Let us now turn to $H^{(1)}$ acting on $|\psi_1\rangle$. We first note that $B_{r_1}|\psi_2\rangle$
has one pair so that closure relation for one-pair
subspace leads to
\begin{equation}
H^{(1)}|\psi_1\rangle=\sum_{r_1,r_2}E_{r_1}B_{r_1}^\dag B_{r_2}^\dag|v
\rangle\langle v|B_{r_2} B_{r_1}|\psi_1\rangle\ ,
\end{equation}
in which we can remove projector $|v\rangle\langle v|$ since $B_{r_2}
B_{r_1}|\psi_1\rangle$ has zero pair.

This readily shows that $H^{(2)}$, such that $H\psi_2\rangle=\left(
H^{(1)}+H^{(2)}\right)|\psi_2\rangle$, can be identified with
\begin{equation}
H^{(2)}=-\frac{1}{2}\sum_{r_1,r_2}E_{r_1}B_{r_1}^\dag B_{r_2}^\dag B_{r_2}
B_{r_1}+\frac{1}{4}\sum_{r'_1,r'_2,r_1,r_2}\xi\left(^{r'_2\ r_2}_{r'_1\
r_1}\right)B_{r'_1}^\dag B_{r'_2}^\dag B_{r_2} B_{r_1}\ .
\end{equation}

\subsection{Derivation of $H^{(3)}$}

To grasp how series $H$ is constructed, let us calculate one more
$H^{(n)}$ explicitly. From closure relation for 3-pair states, we find
\begin{equation}
H|\psi_3\rangle=\left(\frac{1}{3!}\right)^2\sum_{\{r\}}HB_{r_1}^\dag
B_{r_2}^\dag B_{r_3}^\dag|v\rangle\langle v|B_{r_3}B_{r_2}B_{r_1}|\psi_3
\rangle\ .
\end{equation}
We then use Eqs.(1.10,11) to find
\begin{eqnarray}
HB_{r_1}^\dag B_{r_2}^\dag B_{r_3}^\dag|v\rangle=(E_{r_1}+E_{r_2}+E_{r_3})
B_{r_1}^\dag B_{r_2}^\dag B_{r_3}^\dag|v\rangle+\sum_{s,t}B_s^\dag
B_t^\dag \left[\xi\left(^{t\ r_2}_{s\
r_1}\right)B_{r_3}^\dag\right.\nonumber\\
\left.\xi\left(^{t\ r_3}_{s\ r_2}\right)B_{r_1}^\dag+
\xi\left(^{t\ r_1}_{s\ r_3}\right)B_{r_2}^\dag\right]\ .
\end{eqnarray}
So that, if we insert this result into Eq.(3.15), relabel bold
indices and remove projector $|v\rangle\langle v|$, we end with
\begin{eqnarray}
H|\psi_3\rangle=\left[\frac{3}{(3!)^2}\sum_{\{r\}}E_{r_1}B_{r_1}^\dag
B_{r_2}^\dag B_{r_3}^\dag B_{r_3}B_{r_2}B_{r_1}
\right.\hspace{3cm}\nonumber\\
\left.+\frac{C_3^2}{(3!)^2}\sum_{r'_1,r'_2,\{r\}}\xi\left(^{r'_2\ r_2}
_{r'_1\ r_1}\right)B_{r'_1}^\dag B_{r'_2}^\dag B_{r_3}^\dag B_{r_3}
B_{r_2} B_{r_1}\right]|\psi_3\rangle\ ,
\end{eqnarray}
where $C_N^2=N(N-1)/2$ is the number of ways we can choose 2 excitons among
$N$. This makes $C_3^2=3$.

We now turn to $\left(H^{(1)}+H^{(2)}\right)|\psi_3\rangle$ that we
calculate by injecting closure relations for 2-pair states in front of
$B_{r_1}|\psi_3\rangle$ and for one-pair states in front of $B_{r_1}B_{r_2}
|\psi_3\rangle$. By collecting all these results, we find that $H^{(3)}$
such that $H|\psi_3\rangle=\left(H^{(1)}+H^{(2)}+H^{(3)}\right)|\psi_3
\rangle$ can be identified with
\begin{eqnarray}
H^{(3)}=\frac{1}{3}\sum_{\{r\}}E_{r_1}B_{r_1}^\dag
B_{r_2}^\dag B_{r_3}^\dag
B_{r_3}B_{r_2}B_{r_1}-\frac{1}{6}\sum_{r'_1,r'_2,\{r\}}\xi\left(^{r'_2\
r_2}_{r'_1\ r_1}\right)B_{r'_1}^\dag B_{r'_2}^\dag B_{r_3}^\dag
B_{r_3}B_{r_2}B_{r_1}\ .
\end{eqnarray}

\subsection{Derivation of $H^{(n)}$}

The above results lead us to think that operator $H^{(n)}$ can be
written as
\begin{eqnarray}
H^{(n)}=\alpha_n\,\sum_{\{r\}}E_{r_1}B_{r_1}^\dag\ldots B_{r_n}^\dag
B_{r_n}\ldots B_{r_1}\hspace{5cm}\nonumber\\
+\beta_n\,\sum_{r'_1,r'_2,\{r\}}\xi\left(^{r'_2\
r_2}_{r'_1\ r_1}\right)B_{r'_1}^\dag B_{r'_2}^\dag B_{r_3}^\dag\ldots
B_{r_n}^\dag B_{r_n}\ldots B_{r_1}\ ,
\end{eqnarray}
with $\alpha_n=-2\beta_n=\gamma_n$ for
$n>1$, with $\gamma_n$ being the prefactor appearing in $D_{mi}$ series (see Eq.(2.23)), while $(\alpha_1=1,\beta_1=0)$ for $n=1$.

In order to show this nicely simple result, we are going to determine the
recursion relations which exist between
$\alpha_n$'s and between $\beta_n$'s. For that, we follow the
procedure we have previously used, namely, we first insert closure relation
for the $N$-pair states in front of $|\psi_N\rangle$. This leads to
\begin{equation}
H|\psi_N\rangle=\frac{1}{(N!)^2}\sum_{\{r\}}HB_{r_1}^\dag\ldots B_{r_N}
^\dag|v\rangle\langle v|B_{r_N}\ldots B_{r_1}|\psi_N\rangle\ .
\end{equation}
We then calculate $H$ acting on $N$ excitons through Eq.(1.10). This leads to 
\begin{equation}
HB_{r_1}^\dag\ldots B_{r_N}^\dag|v\rangle=(B_{r_1}^\dag H+E_{r_1}B_{r_1}
^\dag+V_{r_1}^\dag)B_{r_2}^\dag\ldots B_{r_N}^\dag|v\rangle\ .
\end{equation}
By using Eq.(1.11), we find
\begin{eqnarray}
V_{r_1}^\dag B_{r_2}^\dag\ldots
B_{r_N}^\dag|v\rangle&=&\left([V_{r_1}^\dag, B_{r_2}^\dag]+B_{r_2}^\dag
V_{r_1}^\dag\right)B_{r_3}^\dag\ldots B_{r_N} ^\dag|v\rangle\nonumber\\
&=&\sum_{r'_1,r'_2}\xi\left(^{r'_2\
r_2}_{r'_1\ r_1}\right)B_{r'_1}^\dag B_{r'_2}^\dag B_{r_3}^\dag\ldots B_{r_N}
^\dag|v\rangle + B_{r_2}^\dag V_{r_1}^\dag B_{r_3}^\dag\ldots B_{r_N}
^\dag|v\rangle\ .
\end{eqnarray}
We iterate the procedure to end with
\begin{eqnarray}
HB_{r_1}^\dag\ldots B_{r_N}^\dag|v\rangle=(E_{r_1}+\cdots+E_{r_N})
B_{r_1}^\dag\ldots B_{r_N}^\dag|v\rangle\hspace{5cm}\nonumber\\
+\left\{\sum_{r'_1,r'_2}\xi\left(^{r'_2\ r_2}_{r'_1\ r_1}\right)B_{r'_1}^\dag
B_{r'_2}^\dag B_{r_3}^\dag\ldots B_{r_N} ^\dag|v\rangle\ +\
\mathrm{permutations}\right\}\ ,
\end{eqnarray}
the total number of these $\xi$ terms being the number of ways $C_N^2$ we can choose among $N$,
the two excitons having direct Coulomb process.

If we now relabel bold indices and remove projector
$|v\rangle\langle v|$, we end with
\begin{eqnarray}
H|\psi_N\rangle=\frac{N}{(N!)^2}\sum_{\{r\}}E_{r_1}B_{r_1}^\dag\ldots
B_{r_N}^\dag B_{r_N}\ldots B_{r_1}|\psi_N\rangle\hspace{4cm}\nonumber\\
+\frac{C_N^2}{(N!)^2}\sum_{r'_1,r'_2,\{r\}}
\xi\left(^{r'_2\ r_2}_{r'_1\ r_1}\right)B_{r'_1}^\dag
B_{r'_2}^\dag B_{r_3}^\dag\ldots B_{r_N} ^\dag B_{r_N}\ldots B_{r_1}
|\psi_N\rangle\ .
\end{eqnarray}

We now turn to $H^{(n)}$ acting on $|\psi_N\rangle$ and assume that its general form is indeed given by Eq.(3.19). Since state
$B_{r_n}\ldots B_{r_1}|\psi_N\rangle$ has $(N-n)$ pairs, we get, by
injecting closure relation for $(N-n)$-pair subspace,
\begin{eqnarray}
H^{(n)}|\psi_N\rangle=\frac{1}{((N-n)!)^2}\sum_{\{r\}}\left[\alpha_nE_{r_1}B_{r_1}
^\dag B_{r_2}^\dag+\beta_n\sum_{r'_1,r'_2}
\xi\left(^{r'_2\ r_2}_{r'_1\ r_1}\right)B_{r'_1}^\dag
B_{r'_2}^\dag\right]\nonumber\\
\times\ B_{r_3}^\dag\ldots B_{r_n}^\dag B_{r_{n+1}}^\dag\ldots B_{r_N}
^\dag|v\rangle\langle v|B_{r_N}\ldots B_{r_1}|\psi_N\rangle\ .
\end{eqnarray}
We then remove projector $|v\rangle\langle v|$ as usual. By
collecting all these results, we find that operator $H^{(n)}$ defined
through Eq.(3.6) has the form (3.19) provided that $\alpha_n$'s
and $\beta_n$'s are linked by
\begin{equation}
\alpha_N=\frac{N}{(N!)^2}-\sum_{n=1}^{N-1}\frac{\alpha_n}{((N-n)!)^2}\ ,
\end{equation}
\begin{equation}
\beta_N=\frac{C_N^2}{(N!)^2}-\sum_{n=2}^{N-1}\frac{\beta_n}{((N-n)!)^2}\
,
\end{equation}
with $\alpha_1=1$ and $\beta_1=0$, due to Eq.(3.9), while $\beta_2=1/4$,
due to Eq.(3.14). By comparing Eqs.(2.23) and (3.26), we readily see that
$\alpha_N=\gamma_N$. In order to determine $\beta_N$, we first note that
the recursion relation for $\alpha_N$ also reads
\begin{eqnarray}
\alpha_N&=& \frac{N}{(N!)^2}-\frac{1}{((N-1)!)^2}-\sum_{n=2}^{N-1}
\frac{\alpha_n}{((N-n)!)^2}\nonumber\\
&=& -\frac{N(N-1)}{(N!)^2}-\sum_{n=2}^{N-1}\frac{\alpha_n}{((N-n)!)^2}\ .
\end{eqnarray}
Since $C_N^2=N(N-1)/2$, this equation is nothing but the recursion
relation for $\beta_N$ provided that we replace $\alpha_N$ by
$(-2\beta_N)$ for any $N\geq 2$. Consequently, we end with 
\begin{equation}
\gamma_N=\alpha_N=-2\beta_N \ \ \mathrm{for}\ \ N\geq 2\ ,
\end{equation}
while $\alpha_1=\gamma_1=1$ and $\beta_1=0$, in agreement with our
original guess.

\section{Discussion}

\subsection{Summary of the results}

The above results lead us to write deviation-from-boson operator $D_{mi}$ of two composite excitons defined as
\begin{equation}
[B_m,B_i^\dag]=\delta_{m,i}-D_{mi}\ ,
\end{equation}
through an
infinite series of exciton-operator products, according to
\begin{equation}
D_{mi}=\sum_{r',r}\left[\lambda\left(^{r'\ \,r}_{m\ i}\right)+
\lambda\left(^{m\ r}_{r'\ \,i}\right)\right]B_{r'}^\dag\left(1+\sum_{n=2}^\infty\gamma_nP_n\right)
B_r\ ,
\end{equation}
\begin{equation}
P_n=\sum_{\{j\}}B_{j_1}^\dag\ldots B_{j_{n-1}}
^\dag B_{j_{n-1}}\ldots B_{j_1}\ .
\end{equation}

$\lambda\left(^{r'\ \,r}_{m\ i}\right)$ is the Pauli scattering for
carrier exchanges between ``in'' excitons $(i, r)$ leading to ``out''
excitons $(m,r')$, with excitons $(m,i)$ having same
electron. Electron-hole symmetry is restored through the fact
that, in the second term of Eq.(4.2), namely, $\lambda\left(^{m\ r}_{r'\ \,i}\right)$, excitons $(m,i)$
have same hole (see Fig.1). 

$\gamma_n$'s are numerical prefactors
which obey the recursion relation
\begin{equation}
\gamma_N=\frac{N}{(N!)^2}-\sum_{n=1}^{N-1}\frac{\gamma_n}{((N-n)!)^2}\ ,
\end{equation}
with $\gamma_1=1$; so that $\gamma_2=-1/2$, $\gamma_3=1/3$, and so on\ldots, with $\gamma_N$ going to zero for increasing $N$.

In the same way, the system Hamiltonian, when acting on
electron-hole-pair states, can be written as an infinite series of
exciton-operator products, according to
\begin{eqnarray}
H=\sum_{r}E_{r}B_{r}^\dag \left(1+\sum_{n=2}^\infty\gamma_nP_n\right)B_{r}\hspace{8cm}\nonumber\\
+\frac{1}{2}
\sum_{r_1,r_2,r'_1,r'_2}\xi\left(^{r'_2\ r_2}_{r'_1\ r_1}\right)B_{r'_1}
^\dag B_{r'_2}^\dag \left(\frac{1}{2}-\sum_{n=3}^\infty\gamma_nP_{n-1}\right)B_{r_2}B_{r_1}\ .
\end{eqnarray}

Let us again stress that, since there are two ways to form two excitons
out of two electron-hole pairs, any product $B^\dag B^\dag$ can be
written as a sum of $B^\dag B^\dag$ according to Eq.(2.4). Consequently,
it is always possible to rewrite sums appearing in $D_{mi}$ and $H$
in various different ways, Eqs.(4.2-5) being
the simplest ones.

\subsection{Comparison with Mukamel and coworkers' results}

In their works, Mukamel and coworkers use free-pair operators
$\hat{B}_{m}^\dag=c_{m_1}^\dag d_{m_2}^\dag$, where $c_{m_1}^\dag$ creates electron on site $m_1$ while $d_{m_2}^\dag$ creates hole on site $m_2$. The fact that they use sites while we here use momenta (see Eq.(3.2)) is basically unimportant. They however keep the possibility for electrons and holes of these free pairs to differ from free
Hamiltonian eigenstates. This is why they have nondiagonal contributions in the one-body part of
their Hamiltonian,
\begin{equation}
H_0=\sum_{m_1,n_1}t^{(1)}_{m_1n_1}c_{m_1}^\dag
c_{n_1}+\sum_{m_2,n_2}t^{(2)}_{m_2n_2}d_{m_2}^\dag d_{n_2}\ .
\end{equation}
As these free-pair states are not physically relevant states to describe a
set of $N$ interacting pairs, Mukamel and coworkers only reach the two
first terms of $H$ series, namely, $H^{(1)}$ and $H^{(2)}$, and the first term of $D_{mi}$ series, their
results reading already as rather complicated (see Eq.(18) of ref.[10]). To make precise link with their work, we are going to recover their results from our compact forms.

As electron-hole states used by Mukamel and coworkers form
complete set, we can expand exciton operators $B_r^\dag$ in terms of
electron-hole operators $\hat{B}_n^\dag$, according to
\begin{equation}
B_r^\dag=\sum_n\hat{B}_n^\dag\langle\hat{n}|r\rangle\ ,
\end{equation}
where $|r\rangle=B_r^\dag|v\rangle$, while $|\hat{n}\rangle=\hat{B}_n^\dag|v
\rangle=c_{n_1}^\dag d_{n_2}^\dag|v\rangle$.

Using this expansion (4.7), we see that the first term of the $H$ series we have obtained, also reads
\begin{eqnarray}
H^{(1)}&=&\sum_{r_1}E_{r_1}B_{r_1}^\dag B_{r_1}\nonumber\\
&=&\sum_{m,n}h_{mn}\hat{B}_m^\dag\hat{B}_n\ ,
\end{eqnarray}
where prefactor $h_{mn}$ is nothing but
\begin{equation}
h_{mn}=\sum_{r_1}E_{r_1}\langle\hat{m}|r_1\rangle\langle r_1|\hat{n}\rangle
=\langle\hat{m}|H|\hat{n}\rangle\ ,
\end{equation}
since $H|r_1\rangle=E_{r_1}|r_1\rangle$. If we now introduce the two-body part of the Hamiltonian as written in Eq.(16) of ref.[10], namely,
\begin{equation}
H_c=\frac{1}{2}\sum V^{(1)}_{m_1n_1j_1k_1}c_{m_1}^\dag c_{n_1}^\dag c_{j_1}c_{k_1}
+\frac{1}{2}\sum V^{(2)}_{m_2n_2j_2k_2}d_{m_2}^\dag d_{n_2}^\dag d_{j_2}d_{k_2}
-\sum W_{m_1n_2j_1k_2}c_{m_1}^\dag d_{n_2}^\dag d_{j_2}c_{k_1}\ ,
\end{equation}
we see that, for $H=H_0+H_c$ with $H_0$ given in Eq.(4.6), prefactor  $h_{mn}$ defined in Eq.(4.9) splits as
\begin{eqnarray}
h_{mn}&=&h_{mn}^{(0)}-W_{m_1m_2n_1n_2}\nonumber\\
h_{mn}^{(0)}&=&t_{m_1n_1}^{(1)}\delta_{m_2,n_2}+t_{m_2n_2}^{(2)}
\delta_{m_1,n_1}\ ,
\end{eqnarray}
in agreement with the result obtained by Mukamel and coworkers for the first term of $H$ expansion.

We now turn to $H^{(2)}$. In view of Eq.(3.14), $H^{(2)}$
splits as $H^{(2)}=H_E^{(2)}+H_{\xi}^{(2)}$, where $H_E^{(2)}$ depends on exciton energy $E_r$ while $H_{\xi}^{(2)}$ depends on exciton scattering $\xi$. Before going further, let us note that, since 
$H_E^{(2)}$ contains
exciton energy $E_{r}$, this term, by construction, contains
a part of electron-hole Coulomb interaction, namely, the one acting \emph{inside one
exciton}. On the other hand, as $H_{\xi}^{(2)}$ reads in terms of direct Coulomb
scattering $\xi\left(^{r'_2\ r_2}_{r'_1\ r_1}\right)$ \emph{between the two excitons $(r_1,r_2)$}, it contains Coulomb scattering resulting
from electron-electron and hole-hole interactions as well as from electron-hole
interaction \emph{between} excitons $r_1$ and $r_2$ (see Fig.3(a)). Since electron-hole interaction already appears in the first-order term $H^{(1)}$ through $W_{m_1m_2n_1n_2}$ in $h_{mn}$ (see Eq.(4.11)), these two electron-hole contributions of $H^{(2)}$ must somehow cancel, as we now show. 

If we symmetrize $H_E^{(2)}$ and write exciton operators in terms of free
pairs according to Eq.(4.7), we find
\begin{eqnarray}
H_E^{(2)}&=&-\frac{1}{4}\sum_{r_1,r_2}(E_{r_1}+E_{r_2})B_{r_1}^\dag
B_{r_2} ^\dag B_{r_2}B_{r_1}\nonumber\\
&=&-\frac{1}{4}\sum_{m,n,j,k}\hat{B}_m^\dag
\hat{B}_n^\dag\hat{B}_j\hat{B}_k\left[\sum_{r_1}\langle\hat{m}|H|r_1\rangle
\langle r_1|\hat{k}\rangle\sum_{r_2}\langle\hat{n}|r_2\rangle\langle r_2|\hat{j}\rangle
+\ (1\leftrightarrow 2)\right]\ .
\end{eqnarray}
Using Eq.(4.9) and orthogonality of free-pair states, we readily
find
\begin{equation}
H_E^{(2)}=-\frac{1}{4}\sum_{m,n,j,k}\hat{B}_m^\dag
\hat{B}_n^\dag\hat{B}_j\hat{B}_k(h_{mk}\delta_{n,j}+h_{nj}\delta_{m,k})\ ,
\end{equation}
with $h_{mn}$ given in Eq.(4.11).

If we now consider the part of $H_{\xi}^{(2)}$ coming from Coulomb
scattering $\xi$ \emph{between} excitons, we can rewrite it, using again
Eq.(4.7), as
\begin{eqnarray}
H_{\xi}^{(2)}&=&\frac{1}{4}\sum_{r_1,r_2,r'_1,r'_2}\xi\left(^{r'_2\ r_2}
_{r'_1\ r_1}\right)B_{r'_1}^\dag B_{r'_2}^\dag B_{r_2}B_{r_1}\nonumber\\
&=&\frac{1}{4}\sum_{m,n,j,k}\hat{B}_m^\dag
\hat{B}_n^\dag\hat{B}_j\hat{B}_k\sum_{r_1,r_2,r'_1,r'_2}\xi
\left(^{r'_2\ r_2}_{r'_1\ r_1}\right)\langle\hat{m}|r'_1\rangle\langle\hat{n}|r'_2
\rangle\langle r_2|\hat{j}\rangle\langle r_1|\hat{k}\rangle\ .
\end{eqnarray}
The sum over $r$'s is readily obtained from diagrams of Fig.3(b) in
terms of interactions $V^{(1)}$ between electrons, $V^{(2)}$ between
holes and $W$ between electrons and holes. It reduces to
\begin{equation}
\left\{V_{m_1n_1j_1k_1}^{(1)}\delta_{m_2,k_2}\,\delta_{n_2,j_2}-
W_{m_1n_2j_2k_1}\delta_{m_2,k_2}\,\delta_{n_1,j_1}\right\}+
\{1\leftrightarrow 2\}\ .
\end{equation}

If we now collect the two parts of $H^{(2)}$, we can rewrite it as
\begin{equation}
H^{(2)}=\sum_{m,n,j,k}U_{mnjk}\hat{B}_m^\dag
\hat{B}_n^\dag\hat{B}_j\hat{B}_k+Z\ ,
\end{equation}
where $U_{mnjk}$ is just the prefactor obtained by Mukamel and coworkers in Eq.(18) of ref.[10]. 
Operator $Z$ contains all electron-hole contributions. Its precise value reads
\begin{eqnarray}
Z=\sum_{m,n,j,k}\hat{B}_m^\dag\hat{B}_n^\dag\hat{B}_j\hat{B}_k\left[
\frac{1}{4}\left(W_{m_1m_2k_1k_2}\delta_{n,j}+W_{n_1n_2j_1j_2}\delta_{m,k}
\right)\right.\nonumber\\
\left.-\frac{1}{4}\left(W_{m_1n_2j_2k_1}\delta_{m_2,k_2}\delta_{n_1,j_1}
+W_{n_1m_2k_2j_1}\delta_{m_1,k_1}\delta_{n_2,j_2}\right)\right]\ .
\end{eqnarray}
In order for Eq.(4.16) to agree with the expression of $H^{(2)}$ obtained by Mukamel and
coworkers, operator $Z$ must reduce to zero. This is actually
true, as shown by noting that
\begin{equation}
\hat{B}_m^\dag\hat{B}_n^\dag\hat{B}_j\hat{B}_k=c_{m_1}^\dag d_{m_2}^\dag c_{n_1}^\dag d_{n_2}^\dag d_{j_2}c_{j_1}d_{k_2}c_{k_1}=
c_{m_1}^\dag(-d_{n_2}^\dag
c_{n_1}^\dag d_{m_2}^\dag)(-d_{j_2}c_{k_1}d_{k_2})c_{j_1}\ ,
\end{equation}
and by exchanging bold indices $(m_2\leftrightarrow n_2)$ and 
$(j_2\leftrightarrow k_2)$ in the sums appearing in $Z$. This explicitly shows that electron-hole interaction does not appear in $H^{(2)}$ as reasonable since, due to Eq.(3.3), $V_{eh}$ can be exactly written in terms of $B_i^\dag B_j$, or $\hat{B}_m^\dag \hat{B}_n$.

We thus conclude that expressions of $H^{(1)}$ and $H^{(2)}$ given by
Mukamel and coworkers agree with our compact form of $H^{(n)}$. 
As the exciton operators we here use are physically relevant operators for interacting
electron-hole pairs, we have been able to write the whole infinite
series for $H$ in a compact form, in terms of these operators. Let us however stress that, even with this infinite
series now known, it is far simpler to calculate $H|\psi_N\rangle$
through the commutators $[H,B_i^{\dag N}]$ and $[V_i^\dag,B_j^{\dag N}]$,
given in Eqs.(1.6,7), than through this $H^{(n)}$ series, mostly when the
state $|\psi_N\rangle$ of interest has many identical excitons, as in
usual physically relevant situations.

\subsection{Possible use of series expansion for $H$}

The procedure proposed by Mukamel and coworkers is definitely not a
bosonization procedure, since \emph{exact} deviation-from-boson
operators are \emph{a priori} kept through their expansion as a series of
pair operators. We can however be tempted by comparing prefactors obtained in this expansion of the system Hamiltonian $H$ in terms of exciton operators, with effective
scatterings produced by bosonization.

When truncated to its one and two-body terms, the effective
Hamiltonian for bosonized excitons reads as
\begin{equation}
\bar{H}=\sum_iE_i\bar{B}_i^\dag\bar{B}_i+\frac{1}{2}\sum_{mnij}\bar{\xi}\left(
^{n\ \,j}_{m\ i}\right)\bar{B}_m^\dag\bar{B}_n^\dag\bar{B}_i\bar{B}_j\ ,
\end{equation}
with $[\bar{B}_m,\bar{B}_i^\dag]=\delta_{m,i}$ for elementary bosons. We
see that the prefactor of the first term of $\bar{H}$ is nothing but the one of $H^{(1)}$. If we now
consider $H^{(2)}$ given in Eq.(3.14), we can rewrite it as
\begin{equation}
H^{(2)}=\frac{1}{4}\sum_{r_1,r_2,r'_1,r'_2}\left[\xi
\left(^{r'_2\ r_2}_{r'_1\ r_1}\right)-(E_{r_1}+E_{r_2})(\delta_{r'_1,r_1}
+\delta_{r'_2,r_2})\right]B_{r'_1}^\dag B_{r'_2}^\dag B_{r_2}B_{r_1}\ .
\end{equation}
Due to the two ways to form two excitons out of two electron-hole pairs
which lead to Eq.(2.4), we get from this equation used for $B^\dag
B^\dag$ or $BB$,
\begin{eqnarray}
\sum_{r_1,r_2,r'_1,r'_2}\xi\left(^{r'_2\ r_2}_{r'_1\ r_1}\right)
B_{r'_1}^\dag B_{r'_2}^\dag B_{r_2}B_{r_1}&=&-
\sum_{r_1,r_2,r'_1,r'_2}\xi^{\mathrm{in}}\left(^{r'_2\ r_2}_{r'_1\
r_1}\right) B_{r'_1}^\dag B_{r'_2}^\dag B_{r_2}B_{r_1}\nonumber\\ &=& -
\sum_{r_1,r_2,r'_1,r'_2}\xi^{\mathrm{out}}\left(^{r'_2\ r_2}_{r'_1\
r_1}\right) B_{r'_1}^\dag B_{r'_2}^\dag B_{r_2}B_{r_1}\ ,
\end{eqnarray}
where $\xi^\mathrm{in}$ and $\xi^\mathrm{out}$, shown in Fig.3(c,d), are
defined as
\begin{eqnarray}
\xi^\mathrm{in}\left(^{r'_2\ r_2}_{r'_1\ r_1}\right)&=&\sum_{p_1,p_2}
\lambda\left(^{r'_2\ p_2}_{r'_1\ p_1}\right)
\xi\left(^{p_2\ r_2}_{p_1\ r_1}\right)\ ,
\\
\xi^\mathrm{out}\left(^{r'_2\ r_2}_{r'_1\ r_1}\right)&=&\sum_{p_1,p_2}
\xi\left(^{r'_2\ p_2}_{r'_1\ p_1}\right)
\lambda\left(^{p_2\ r_2}_{p_1\ r_1}\right)\ .
\end{eqnarray}
This shows that in Eq.(4.20), $\xi$ can be replaced by
$(-\xi^\mathrm{in})$ or $(-\xi^\mathrm{out})$, or even by 
$(a\xi-b\xi^\mathrm{in}-c\xi^\mathrm{out})$ with $a+b+c=1$.

If we now turn to the $E$ part of $H^{(2)}$, the same Eq.(2.4) used for
$B^\dag B^\dag$ or $BB$ leads to
\begin{eqnarray}
\sum_{r_1,r_2}(E_{r_1}+E_{r_2})B_{r_1}^\dag B_{r_2}^\dag B_{r_2}B_{r_1}&=&
-\sum_{r_1,r_2,r'_1,r'_2}(E_{r_1}+E_{r_2})\lambda
\left(^{r'_2\ r_2}_{r'_1\ r_1}\right)B_{r'_1}^\dag B_{r'_2}^\dag
B_{r_2}B_{r_1}\nonumber\\
&=& -\sum_{r_1,r_2,r'_1,r'_2}(E_{r'_1}+E_{r'_2})\lambda
\left(^{r'_2\ r_2}_{r'_1\ r_1}\right)B_{r'_1}^\dag B_{r'_2}^\dag
B_{r_2}B_{r_1}\ ,
\end{eqnarray}
so that $E$ prefactor in $H^{(2)}$ gives rise to two-body scattering between excitons in
$E\lambda$. 

 This shows that the second term of $H$ expansion can also be written as
 \begin{equation}
 H^{(2)}=\frac{1}{2}\sum_{r_1,r_2,r'_1,r'_2}S\left(^{r'_2\ r_2}_{r'_1\ r_1}\right)B_{r'_1}^\dag B_{r'_2}^\dag
B_{r_2}B_{r_1}\ ,
\end{equation}
\begin{equation}
S\left(^{r'_2\ r_2}_{r'_1\ r_1}\right)=\frac{1}{2}\left[a\xi\left(^{r'_2\ r_2}_{r'_1\ r_1}\right)-b\xi^\mathrm{in}
\left(^{r'_2\ r_2}_{r'_1\ r_1}\right)-c\xi^\mathrm{out}\left(^{r'_2\ r_2}_{r'_1\ r_1}\right)-(E_{r'_1}+E_{r'_2})
\lambda\left(^{r'_2\ r_2}_{r'_1\ r_1}\right)\right]\ ,
\end{equation}
with $a+b+c=1$. It is however clear that such a $S\left(^{r'_2\ r_2}_{r'_1\ r_1}\right)$ cannot be used as an effective scattering between two excitons. Indeed, $S$ depends
on energy origin through exciton energy $E_i$ which includes the band gap in the case of
excitons, while it is physically irrelevant to have the band gap entering exciton scattering. Even if we drop these spurious $E\lambda$ terms, this $S\left(^{r'_2\ r_2}_{r'_1\ r_1}\right)$ has problem since its $\xi$ part differs from the effective scattering between bosonized excitons mostly used in the literature, namely $\bar{\xi}\left(^{n\ \,j}_{m\ i}\right)=\xi\left(^{n\ \,j}_{m\ i}\right)-\xi^\mathrm{out}\left(^{n\ \,j}_{m\ i}\right)$, by at least a factor of 1/2, in addition to the fact that effective Hamiltonians with such a $\bar{\xi}$ are not hermitian: Indeed, in order for $\bar{H}$ to be hermitian, we must have $\bar{\xi}\left(^{n\ \,j}_{m\ i}\right)=\bar{a}\xi\left(^{n\ \,j}_{m\ i}\right)-\bar{b}\xi^\mathrm{in}\left(^{n\ \,j}_{m\ i}\right)-\bar{c}\xi^\mathrm{out}\left(^{n\ \,j}_{m\ i}\right)$, with $\bar{a}=\bar{a}^\ast$ and $\bar{b}=\bar{c}^\ast$. With respect to hermiticity, let us recall that Eqs.(4.25,26) are written with  composite-boson operators, not
elementary-boson operators $\bar{B}_r$: This makes Eq.(4.21) correct, \emph{i.e.}, $H^{(2)}$  hermitian, even for $a\neq a^\ast$ and $b\neq c^\ast$.

\section{Conclusion}

In this paper, we revisit the procedure proposed by Mukamel and coworkers to
approach interactions between excitons while keeping their composite
nature exactly, through infinite series of composite-boson operators
for both, the system Hamiltonian and the deviation-from-boson operator
of these composite bosons. While Mukamel and coworkers use
free-electron-hole-pair operators, we here use exciton operators which are physically relevant operators for problems dealing with excitons. This
allows us to write all terms of these two infinite series explicitly. They read in terms of
exciton energies as well as Pauli and interaction scatterings that appear in
the composite-boson many-body theory we have recently constructed. We show that 
the first-order terms found by Mukamel and coworkers agree with our results. However, the necessary handling of these two infinite series for calculations
dealing with $N$ excitons makes this approach far more complicated 
than the ones based on the many-body theory for composite bosons we have proposed and which
only relies on four nicely simple commutators.

\newpage

\begin{figure}[t]
\centerline{\scalebox{0.8}{\includegraphics{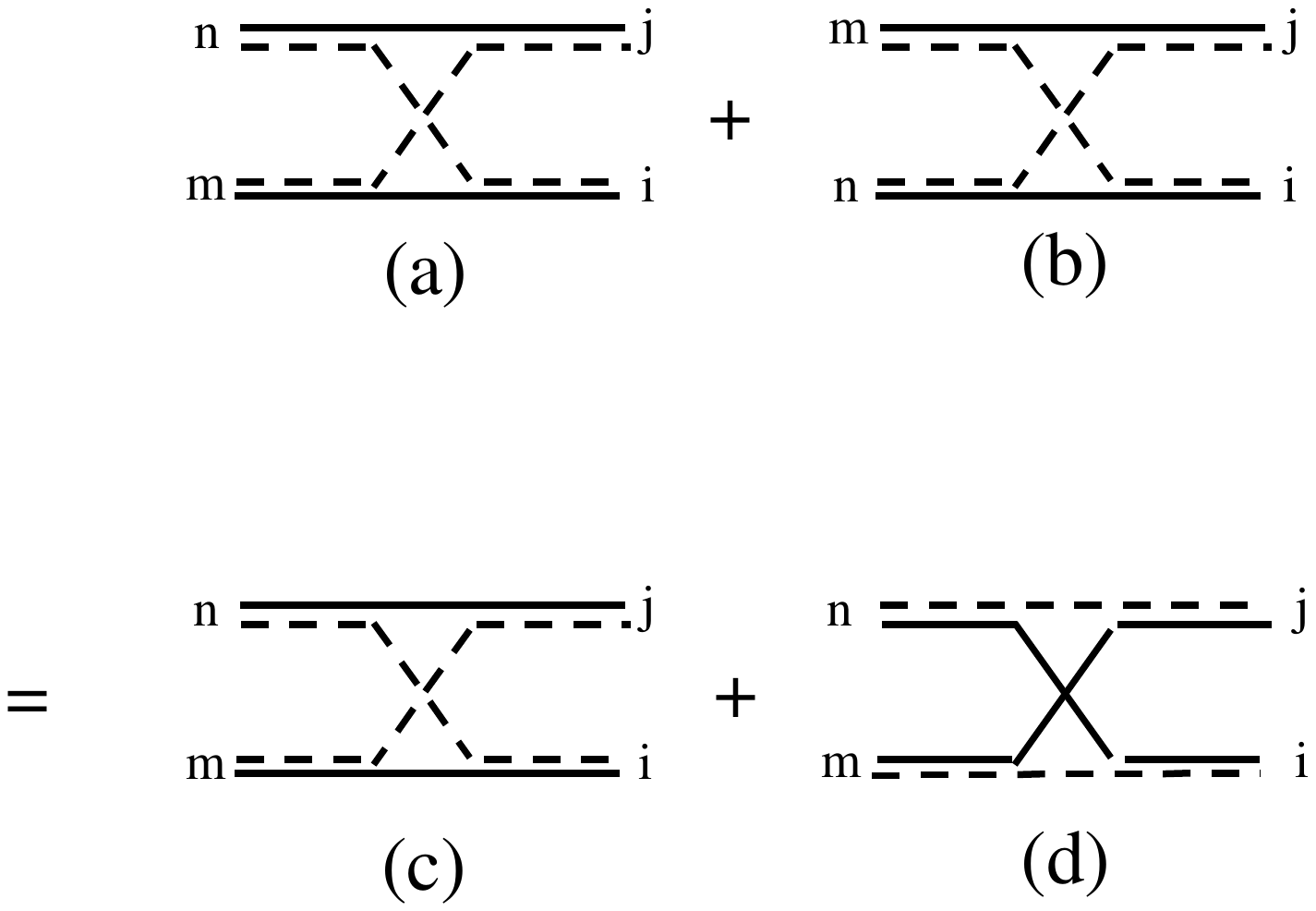}}}
\caption{Shiva diagrams for $\Lambda_{mi}(n,j)$ defined in Eq.(1.12).
$\lambda\left(^{n\ \,j}_{m\ i}\right)$, represented by diagram (a), is
identical to $\lambda_e\left(^{n\ \,j}_{m\ i}\right)$, represented by diagram (c), in which $m$ and $i$
have the same electron. $\lambda\left(^{m\ j}_{n\ \,i}\right)$, represented by diagram (b), is
identical to $\lambda_h\left(^{n\ \,j}_{m\ i}\right)$, represented by diagram (d), in which $m$ and $i$
have the same hole.}
\end{figure}

\newpage

\begin{figure}[t]
\vspace{-3cm}
\centerline{\scalebox{0.8}{\includegraphics{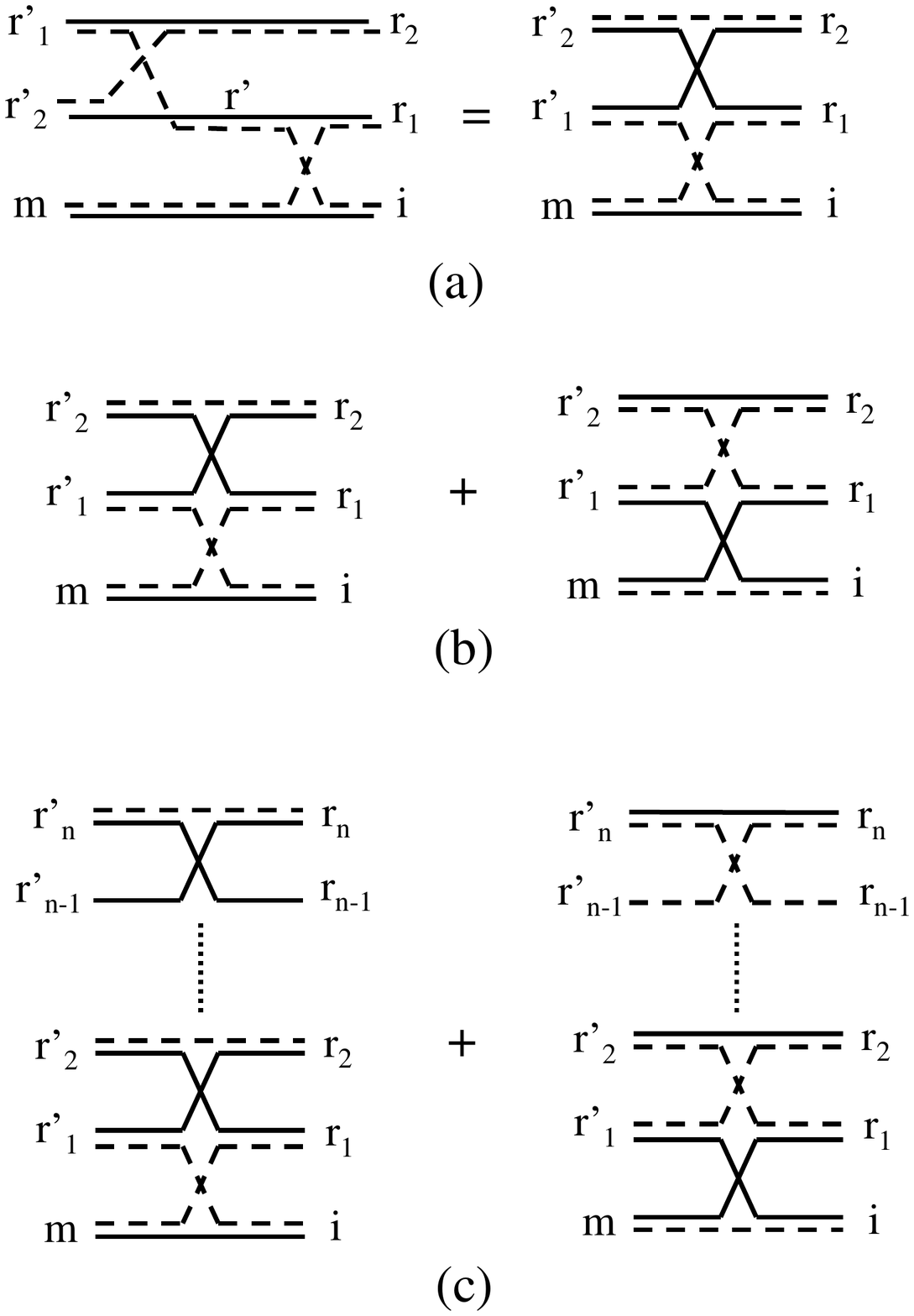}}}
\vspace{-3cm}
\caption{(a) Shiva diagram representation of Eq.(2.27), the summation over the bold index $r'$ being performed readily. (b) Shiva diagrams for the prefactor $\Lambda_{mi}\left(^{r'_2\ r_2}_{r'_1\ r_1}\right)$ appearing in Eq.(2.28). This prefactor corresponds to carrier exchanges between $(i,r_1,r_2)$ leading to $(m,r'_1,r'_2)$, in which the excitons $m$ and $i$ either have the same electron or the same hole. (c) Same as (b) for the prefactor appearing in Eq.(2.29).}
\end{figure}

\newpage

\begin{figure}[t]
\vspace{-3cm}
\hspace{1cm}
\centerline{\scalebox{0.8}{\includegraphics{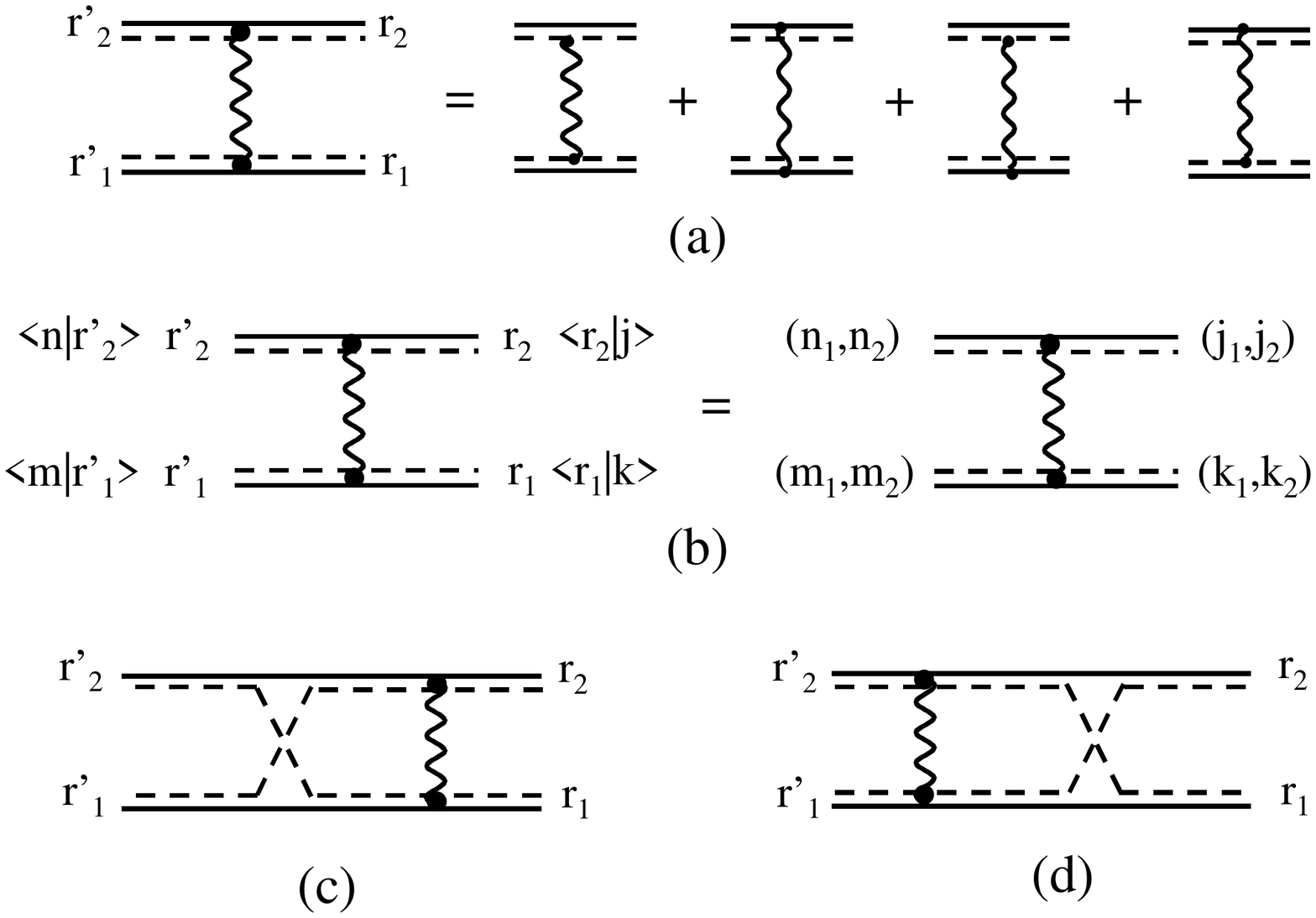}}}
\vspace{-3cm}
\caption{(a) Direct Coulomb scattering $\xi\left(^{r'_2\ r_2}_{r'_1\ r_1}\right)$ between the ``in'' excitons $(r_1,r_2)$ leading to the ``out'' excitons $(r'_1,r'_2)$. This scattering is ``direct'' in the sense that the electron-hole pairs are coupled similarly in the ``in'' and ``out'' states. (b) Diagrammatic representation of the sum over $\{r\}$ appearing in Eq.(4.14) and leading to Eq.(4.15). (c) Shiva diagram representation for the ``in'' exchange Coulomb scattering $\xi^\mathrm{in}\left(^{r'_2\ r_2}_{r'_1\ r_1}\right)$ defined in Eq.(4.22). In this scattering, the Coulomb interactions are between the ``in'' excitons. (d) Shiva diagram representation of the ``out'' exchange Coulomb scattering
$\xi^\mathrm{out}\left(^{r'_2\ r_2}_{r'_1\ r_1}\right)$ defined in Eq.(4.23). In this scattering, the Coulomb interactions are between the ``out'' excitons. }
\end{figure}

\end{document}